# Quantum Connectivity of Space-Time and Gravitationally Induced De-correlation of Entanglement

T.C.Ralph, G.J.Milburn and T.Downes

Department of Physics, University of Queensland, Brisbane 4072, QLD, Australia

(Dated: October 23, 2018)

We discuss an alternative formulation of the problem of quantum optical fields in a curved space-time using localized operators. We contrast the new formulation with the standard approach and find observable differences for entangled states. We propose an experiment in which an entangled pair of optical pulses are propagated through non-uniform gravitational fields and find that the new formulation predicts de-correlation of the optical entanglement under experimentally realistic conditions.

## I. INTRODUCTION

The semi-classical extension of quantum field theory to curved space-times is a well developed theory [1]. However, its applicability is restricted to "well-behaved" metrics. Even then unresolved issues remain, the most famous being the apparent global non-unitarity of the theory implied by Hawking radiation from black holes [2]. Progress in resolving such issues is hampered by a lack of experimental indicators. Typically, situations in which competing approaches make testable predictions involve experimental scenarios far beyond the reach of current technology.

An extreme example of a badly-behaved metric is the wormhole metric introduced by Morris et al [3]. Such a metric allows the existence of time-like curves. Time-like curves allow a particle to follow a trajectory into its own past. Surprisingly, in a model originally introduced by Deutsch [4] and later developed by Bacon [5], it was shown that consistent quantum evolutions can exist in the presence of time-like curves. Although these evolutions appear to be intrinsically non-unitary, one of us has recently shown that an equivalent consistent unitary model can be constructed [6].

The essential physics of this unitary model is that operators describing observables at different points along the particle's geodesic must act on independent Hilbert sub-spaces, and hence commute [6]. This is required because the time-like curve allows different points along the geodesic to interact with each other (see Fig.1). In standard quantum field theory operators at different points along the geodesic are assumed to act on a common sub-space and so in general do not commute. This then raises the question: is it possible to construct a non-standard field theory that contains commutability along the geodesic, but nonetheless reproduces the predictions of standard quantum field theory in flat space? Such a theory might be more generally applicable and could offer a general solution to the problem of non-unitarity. If a non-standard theory of this type can be constructed, an important question to ask is: under what conditions would testable differences between the standard and non-standard approaches arise?

There are additional reasons that cause us to question the standard approach. As Penrose has emphasised [7], there is an apparent conflict between the intrinsic locality of general relativity and the non locality of quantum mechanics. While there seems to be a peaceful co existence between special relativity and quantum non locality, this may not be so easy to maintain in general curved space times. First of all , in both flat and curved space time the propagation of a photon is carried by a phase shift, but the dependance of this phase on frequency can be quite different in curved space time to account for the gravitational red shift. If this phase shift can be accurately measured it will give information on the curvature of the field. While we do not normally think of a single photon pulse as making a measurement on flat space time, that interpretation seems almost inevitable in curved space time. However then we must face the well known difficulty of interfacing a quantum object and a classical field [8]: in the standard theory of quantum fields in curved space time there is no quantum back action on the photon due to gravitational curvature. However a measurement interpretation would require such a back action.

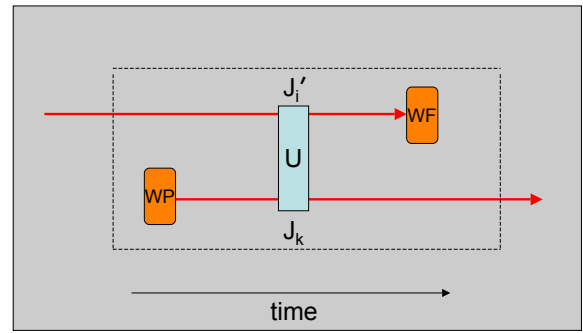

FIG. 1: Representation of the unitary interaction of a quantum system with its past via a wormhole. Consistent evolutions result if operators describing the past ($J_i'$) and future ($J_k$) manifestations of the quantum system act on independent Hilbert sub-spaces. WF is the future mouth of the wormhole, WP is the past mouth.

Secondly, quantum entanglement leads to some strange consequences if the effect of gravitational curvature is purely deterministic. To see this note that two field modes prepared in a squeezed state are entangled in such a way that it is an eigenstate of photon number difference and a near eigenstate of phase sum [9],

$$|\psi\rangle = \sum_n c_n |n\rangle_a |n\rangle_b \qquad (1)$$

Any deterministic phase shift on one mode is so tightly correlated with the other that it can be attributed to either. Suppose one mode passes through a region of curved space time, undergoing a complicated, but deterministic, phase shift, while the other passes only through a flat space-time. The state would change accordingly as

$$|\psi\rangle = \sum_n c_n e^{i\phi(n)} |n\rangle_a |n\rangle_b \qquad (2)$$

It is then always possible to remove the phase shift from the entangled state entirely by operations on the photon in the flat space time region, provided that phase shift is deterministic and completely known by all observers in principle. This would appear to conflict with the locality of general relativity.

In this paper we make an initial attempt to construct a non-standard theory and explore its properties. We restrict ourselves to the quantized electro-magnetic field in 2 dimensions, 1 space and 1 time, and consider the properties of our non-standard theory in Minkowski and Schwartzschild metrics. We show that for inertial observers in flat, Minkowski space, all expectation values of the non-standard theory agree with those of the standard approach. In contrast, we predict a testable difference between the two theories for entangled states in curved Schwarzschild space.

We begin in the next section by reviewing the standard field theory approach to quantum optics in terms of mode operators in flat and curved space. We explicitly consider classical and quantum correlated pairs of modes. In Section III we introduce the generalized version of the theory that allows for commutability along the geodesic. In Section IV we model a specific correlation experiment with the two approaches and find a testable difference in curved space under certain conditions. We conclude with a summary and discussion in Section V.

## II. MODE OPERATORS

The standard approach in quantum optics is to expand the optical fields over a set of modes. Evolution from input to output can be thought of as a rearrangement of the modes and their conjugates, dictated by unitary operators which couple (sometimes non-linearly) the various modes involved together. Calculations can be performed in the Heisenberg picture by re-writing the output mode arrangement (as seen by the detector) in terms of the input modes and taking expectation values over the initial state. Explicitly we can write the mode at the detector, $\hat{a}_m$, as

$$\hat{a}_m = F(\hat{a}_1, \hat{a}_1^\dagger, \hat{a}_2, \hat{a}_2^\dagger ...) \qquad (3)$$

where $\hat{a}_i$ are the various input modes and the function $F$ is determined by the unitaries, $U_i(\hat{a}_1, \hat{a}_1^\dagger, \hat{a}_2, \hat{a}_2^\dagger ...)$, acting between the input and the detector. Hence the expectation value for a photon number measurement is found from

$$n = \langle \psi | \hat{a}_m^\dagger \hat{a}_m | \psi \rangle \qquad (4)$$

where $|\psi\rangle$ is the initial state. In this discussion we will mostly assume this initial state is the vacuum state, $|\psi\rangle = |0\rangle$, defined by $\hat{a}|0\rangle = 0$ for all modes. As such all evolution is carried by the operators. More generally we can consider multi-mode photon number correlations of the form

$$C_{av} = \langle 0 | \Pi_i \hat{a}_{mi}^\dagger \hat{a}_{mi} | 0 \rangle \qquad (5)$$

where each of the detector modes, $a_{mi}$, are given by functions of the same general form as Eq.3.

We will specifically consider two unitaries in our examples. The first is the displacement unitary,

$$\hat{D}(\alpha) = e^{(\hat{a}\alpha^* - \hat{a}^\dagger \alpha)} \qquad (6)$$

whose action on the vacuum state is to produce a coherent state. The Heisenberg evolution for displacement is

$$\hat{D}^\dagger(\alpha) \hat{a} \hat{D}(\alpha) = \hat{a} + \alpha. \qquad (7)$$

The second is the parametric entangling unitary, $\hat{U}(\chi)$,

$$\hat{U}(\chi) = e^{(\chi \hat{a}_1 \hat{a}_2 - \chi^* \hat{a}_1^\dagger \hat{a}_2^\dagger)} \qquad (8)$$

whose action on a pair of vacuum modes is to produce a time-energy entangled state. The Heisenberg evolutions for parametric entanglement are

$$\begin{aligned}
\hat{U}^\dagger(\chi) \hat{a}_1 \hat{U}(\chi) &= Cosh\chi\, \hat{a}_1 + Sinh\chi\, \hat{a}_2^\dagger, \\
\hat{U}^\dagger(\chi) \hat{a}_2 \hat{U}(\chi) &= Cosh\chi\, \hat{a}_2 + Sinh\chi\, \hat{a}_1^\dagger. \qquad (9)
\end{aligned}$$

### A. Mode Operators in Flat Space

Space-time parametrization is introduced via superpositions of frequency modes. For example in terms of plane wave modes we can write the mode annihilation operator for a space-time field in flat space traveling in the positive $x$ direction as

$$\hat{a}(t,x) = \int dk\, G(k)\, e^{ik(x-t+\phi^+)} \hat{a}_k \qquad (10)$$

where we have written $t$ in units of space such that $c = 1$ and hence the optical frequency, $\omega_k = |k|$, for wavenumber, $k$. $G(k)$ is a normalised spectral mode distribution function centred around some positive wave number, $k_0$, and is required to be zero for $k < 0$. The single frequency mode annihilation operators, $\hat{a}_k$, are assumed to

3have the commutator $[\hat{a}_{k_1}, \hat{a}_{k_2}^\dagger] = \delta(k_1 - k_2)$. This leads to the same time commutator

$$[\hat{a}(t, x_1), \hat{a}(t, x_2)^\dagger] = \int dk |G(k)|^2 e^{ik(x_1 - x_2)} \quad (11)$$

that characterizes the emission and/or detection spatial mode shape at some fixed time $t$. The phase factor $\phi^+$ is determined by the choice of boundary conditions.

The equivalent form of Eq.10 for a field traveling in the negative x direction is [10]

$$\hat{a}(t, x) = \int dk\, G(k)\, e^{ik(-x-t+\phi^-)} \hat{a}_k \quad (12)$$

If these two oppositely propagating fields are coupled via a mirror at $x = x_m$ then continuity at the boundary requires $\phi^- - \phi^+ = 2x_m$.

The operator $\hat{a}(t, x)$ and its conjugate can be used to represent states and detection in the usual way. To illustrate this consider the projection operator defined by,

$$\hat{P}(x, t) = \hat{a}_\pm^\dagger(x, t)|0\rangle\langle 0|\hat{a}_\pm(x, t) \quad (13)$$

with support on the one particle sector of Fock space. This describes a non-absorbing single photon detector with respect to the inertial coordinate system $x, t$. The probability that the detector records the result "1" from a field state $|\psi\rangle$ is

$$p_1(x, t) = \langle\psi|\hat{P}(x, t)|\psi\rangle = |\langle 0|\hat{a}_\pm(x, t)|\psi\rangle|^2 \quad (14)$$

In the case of a single photon state, defined

$$|\nu\rangle = \int dk \nu(k) \hat{a}_k^\dagger |0\rangle \quad (15)$$

we find that

$$p_1(x, t) = |(\tilde{G}_\pm * \tilde{\nu}^*)(t - x)|^2 \quad (16)$$

where $\tilde{G}_\pm(x)$ and $\tilde{\nu}(x)$ are the Fourier transforms of $G(k)e^{ik\phi_\pm}$ and $\nu(k)$ respectively. This has the expected form of a convolution of a response function that characterises a detector, $\tilde{G}(x)$, and the response function of an infinite bandwidth detector $\tilde{\nu}(x)$. For the special case of a single photon state, the average number of photons detected is just given by $n(x, t) = \langle\nu|\hat{a}_\pm^\dagger(x, t)\hat{a}_\pm(x, t)|\nu\rangle = p_i(x, t)$.

The generalization of the displacement operator, Eq.6, to space-time modes is

$$\hat{D}(\alpha) = e^{\int dk (H(k) e^{ik(x-t+\phi^c)} \hat{a}_k \alpha_{max}^* - h.c.)} \quad (17)$$

where $H(k)$ and $\phi^c$ describe the spectral structure and phase respectively of the classical pulse producing the interaction, and $\alpha_{max}$ represents the maximum value of the displacement, achieved when there is perfect matching between the classical and quantum modes. The displacement unitary acting on the vacuum state produces a coherent state. The photon number expectation value of this coherent state for detection in the mode represented by Eq.10 is $n(x, t) = \langle\alpha|\hat{a}^\dagger(x, t)\hat{a}(x, t)|\alpha\rangle = |\alpha|^2$, where

$$\alpha = \int dk\, G(k) H(k)^*\, e^{ik(\phi^+ - \phi^c)} \alpha_{max} \quad (18)$$

which again is in the form of a convolution with the detector response. The Heisenberg evolution of the mode is as given by Eq.7, but with $\alpha$ as given by Eq.18.

We consider the following generalization of the parametric unitary, Eq.8,

$$\hat{U}(\chi) = e^{\int\int dk dk' (\chi_{max} H(k) H(k')\, e^{i(k+k')(x-t+\phi^c)} \hat{a}_{1k} \hat{a}_{2k'} - h.c.)} \quad (19)$$

More generally, spectral entanglement is produced by the parametric unitary, leading to a multi-mode output. Here, for simplicity, we are considering a special case in which the crystal and pump parameters are chosen to be such that no spectral entanglement occurs [11]. Even with this restriction the Heisenberg evolution only remains of the form in Eq.9 if the modes being coupled have identical spectral and phase structure. More generally Eq.9 goes to

$$\hat{U}^\dagger(\chi) \hat{a}_1 \hat{U}(\chi) = Cosh\chi_{max}\, \hat{a}_1 + Sinh\chi_1\, \hat{a}_{2c}^\dagger$$
$$\hat{U}^\dagger(\chi) \hat{a}_2 \hat{U}(\chi) = Cosh\chi_{max}\, \hat{a}_2 + Sinh\chi_2\, \hat{a}_{1c}^\dagger \quad (20)$$

where

$$\hat{a}_{1c}(t, x) = \int dk\, H(k)\, e^{ik(x-t+\phi^c)} \hat{a}_{1k}$$
$$\hat{a}_{2c}(t, x) = \int dk\, H(k)\, e^{ik(x-t+\phi^c)} \hat{a}_{2k} \quad (21)$$

and $\chi_1$ characterises the overlap of $\hat{a}_1$ with the classical pump and $\chi_2$ characterises the overlap of $\hat{a}_2$ with the classical pump via

$$\chi_j = \int dk\, G_j(k) H(k)^*\, e^{ik(\phi_j^+ - \phi^c)} \chi_{max} \quad (22)$$

with $j = 1, 2$.

Including space time parametrization we now write the expression for the detector mode (Eq.3) in terms of the input modes evaluated at the initial time $t_i$ such that

$$\hat{a}_m = F(\hat{a}_1(t_i, x_1), \hat{a}_1^\dagger(t_i, x_1), \hat{a}_2(t_i, x_2), \hat{a}_2^\dagger(t_i, x_2)...) \quad (23)$$

Expectation values are then evaluated, as per Eq.5, that depend only on the same time commutators of the field operators and the classical parameters.

### B. Mode Operators in a Schwarzschild Metric

We can generalize the mode operators of the previous section to describe radial propagation close to a massive,

non-spinning body of mass $M$. The Schwarzschild metric in the radial direction for such a body is given by

$$d\tau^2 = (1 - \frac{2M}{r}) \, dt^2 - \frac{dr^2}{(1 - \frac{2M}{r})} \qquad (24)$$

where $t$ is the time interval measured by clocks in a distant inertial frame at rest with respect to the massive body and $r$ is the reduced circumference. The generalization of the mode function of Eq.10 to the metric of Eq.24 is

$$\hat{a}(t,x) = \int dk \; G(k) \; e^{ik(r+2Mln(r)-t+\phi^+)} \hat{a}_k \qquad (25)$$

which can be obtained by solving the 2D EM wave equation in the Schwarzschild metric or more elegantly from the conformal equivalence of the Schwarzschild and Minkowski metrics in 2D under the co-ordinate transformation $r \to r + 2Mln(r), t \to t$ [12], [1]. In the standard approach it is assumed that, to the extent that back action on the metric can be neglected, all the physics is carried by the mode operators.

## III.  EVENT OPERATORS

A feature of the standard approach is that all points along the geodesic of the light ray are equivalent. That is a translation of Eq.10 by $x \to x+d, t \to t+d$ produces no change in the mode operator. Similarly Eq.12 is invariant under the translation $x \to x - d, t \to t + d$. For the Schwarzschild metric, Eq.25, the invariant translation is $r \to r + d, t \to t + d + 2Mln(1 + r/d)$. In effect the mode operator is a global operator describing the entire geodesic. As such, different points along the geodesic act on the same Hilbert sub-space and hence in general do not commute, e.g. $[\hat{a}(t,x), \hat{a}^\dagger(t+d, x+d)] = 1$ (for flat space). As discussed in the introduction, we wish to investigate the effect of introducing an independent, local temporal parametrization of the quantum optical modes that lifts this degeneracy along the geodesic.

We proceed in the flowing way. We first construct the detection mode operator in the standard way, evaluated at the detector, i.e.

$$\hat{a}_m(t_d, x_d) = F(\hat{a}_1(t_d, x_d), \hat{a}_1^\dagger(t_d, x_d), \hat{a}_2(t_d, x_d)...) \qquad (26)$$

We then generalize this detection mode operator to a detection *event* operator by adding a second spectral degree of freedom, $\Omega$, and a distribution, $J(\Omega)$, over this degree of freedom to each of the input modes, such that

$$\bar{a}_m(t_d, x_d) = F(\bar{a}_1(t_d, x_d), \bar{a}_1^\dagger(t_d, x_d), \bar{a}_2(t_d, x_d)...) \qquad (27)$$

where the input event operators have the form

$$\bar{a}_i(x_d, t_d) = \int dk \; G(k) \; e^{ik(x_d - t_d + \phi^+)}$$
$$\times \int d\Omega \; J(\Omega) \; e^{i\Omega(t_d)} \bar{a}_{i,k,\Omega} \qquad (28)$$

and the spectral event operators have the non-zero commutator $[\bar{a}_{i,k,\Omega}, \bar{a}^\dagger_{i,k',\Omega'}] = \delta(k-k')\delta(\Omega - \Omega')$. This localizes the mode operator to the region of the detection event. The detection event is centred on the space-time point $(x_d, t_d)$ with a spatial uncertainty characterized by the variance of $|\tilde{G}(x)|^2$, where $\tilde{G}(x)$ is the Fourier transform of $G(k)$, and a temporal uncertainty characterized by the variance of $|\tilde{J}(t)|^2$, where $\tilde{J}(t)$ is the Fourier transform of $J(\Omega)$. We then propagate the detection event operator back along the geodesics of the input modes to the initial state. The phase of $\Omega$ evolves according to local time along the geodesics. Specifically, propagation of a particular input event operator back along its geodesic to an initial state at space-time $x, t$ gives

$$\bar{a}_i(x, t) = \int dk \; G(k) \; e^{ik(x-t+\phi^+)}$$
$$\times \int d\Omega \; J(\Omega) \; e^{i\Omega(t_d - \tau(t))} \bar{a}_{i,k,\Omega}. \qquad (29)$$

where the parameter $\tau(t)$ records the propagation time between $t_d$ and $t$ as incrementally measured by a set of local observers along the light path of this particular mode, i.e.

$$\tau(t) = \int_t^{t_d} ds \qquad (30)$$

where $ds$ is the propagation time across an incremental local frame. We require that these local frames are all at rest with respect to the detection frame (see Fig.2) where in particular the detection frame is that in which the macroscopic device that generates the measurement results is at rest. This definition of the time interval has the feature that it is an invariant and locally defined quantity. Note also that because of the invariance of the mode operator under translation we have $x - t = x_d - t_d$.

We can now define number correlation expectation values in terms of these event operator (in analogy with Eq. 5) as

$$C_{av} = \langle 0 | \Pi_j \bar{a}^\dagger_{mj} \bar{a}_{mj} | 0 \rangle \qquad (31)$$

where now the initial vacuum state is taken to be the global ground state of the event operator Hilbert space via $\bar{a}|0\rangle = 0$ for all event operators.

For the wormhole metric discussed in the introduction (see Fig.1), $\tau(t)$ will be different for different paths through the interaction. For example, the unitary may couple parts of the field several times through the wormhole, whilst other parts may not pass through the wormhole at all. The $\Omega$ degree of freedom will then distinguish between these paths and allow a consistent solution to be constructed [6]. Note also that the description is now explicitly local. In the following we will ask what the effect is of introducing event operators in Minkowski and Schwarzschild space-times.

## A. Event Operators in Flat Space

For an inertial detection frame in flat space all the local observers along the mode paths are in the same inertial frame (i.e. the detection frame) so from Eq.30, $\tau = t_d - t$, and all the input event operators have the form

$$\bar{a}_i(t,x) = \int dk\, G(k)\, e^{ik(x-t+\phi^+)}$$
$$\times \int d\Omega\, J(\Omega)\, e^{i\Omega t}\bar{a}_{i,k,\Omega}. \quad (32)$$

Notice that the same time commutator for the event operator, Eq.32, is identical to that for the equivalent mode operator, Eq.10, i.e.

$$[\hat{a}(t,x), \hat{a}(t,x')^\dagger] = [\bar{a}(t,x), \bar{a}(t,x')^\dagger] = \int dk |G(k)|^2 e^{ik(x-x')} \quad (33)$$

where we have used the normalization of the $J(\Omega)$ function. Notice also that the generalization to event operators does not change any of the classical parameters. Hence we can conclude that for inertial observers in flat space all expectation values remain the same under the transformation from mode operators to event operators.

## B. Event Operators in a Schwarzschild Metric

The calculation of $\tau$ is not so trivial when we consider curved space. The frames required to calculate $\tau$ for radial propagation in a Schwarzschild Metric are the so-called stationary "shell" frames. The local proper intervals at a shell frame at radius $r$ are given by [13]

$$ds = \sqrt{1 - \frac{2M}{r}}\, dt$$
$$dl = \frac{dr}{\sqrt{1 - \frac{2M}{r}}}. \quad (34)$$

We can rewrite Eq.24 as $d\tau^2 = ds^2 - dl^2$ in the shell frame. For free optical propagation $d\tau = 0$, hence $ds = dl$ (i.e. the speed of light is always found to be $c = 1$ when measured locally). As a result $\int ds = \int dl$ and we find

$$\tau(t) = \int_{r(t)}^{x_d} \frac{dr'}{\sqrt{1 - \frac{2M}{r'}}}. \quad (35)$$

where the reduced circumference, $r(t)$, corresponding to the initial far away time, $t$, can be found from the modal phase relations.

Because of the non-trivial expression for $\tau$ in curved space, in general the same time commutators of the mode operators and event operators will differ. This can lead to observable differences in the expectation values calculated from the two approaches as we show in the following.

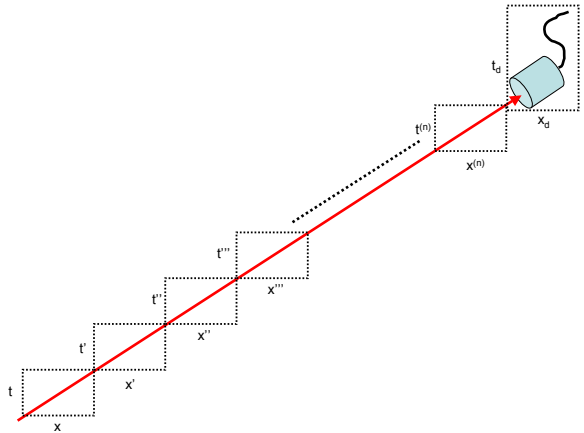

FIG. 2: Representation of local, mutually stationary reference frames for calculating $\tau$.

## IV. A CORRELATION EXPERIMENT IN CURVED SPACE

In the previous two sections we reviewed the standard modal approach to quantum optics in flat space and then generalized this approach to include a speculative additional degree of freedom parameterized by the local propagation time along the mode, $\tau$. We now apply this generalized model to a generic correlation experiment, shown in Fig.3, and allow for space-time curvature. We restrict the problem to two-dimensions, one spatial and time. A source is assumed to populate a pair of orthogonal polarization modes in a correlated way. Initially the polarization modes are spatially degenerate and propagate radially towards a non-spinning massive body. A polarizing beamsplitter located at $x_p$ reflects one polarization mode radially outward whilst the other continues inward until it is also reflected outwards from a mirror located at $x_m$. The modes are observed at time $t_{d1}$ and $t_{d2}$ by detectors situated respectively at $x_{d1}$ and $x_{d2}$ and the measurement results are fed into a correlator, $C$, that multiplies the photo-currents. For this example we assume the detectors are placed far away from the massive body and are at rest with respect to each other, the correlator and the body. Similarly the source is also assumed to be far from, and at rest with respect to the body. The body has a mass $M$, a radius smaller than $x_m$ and is centred at the origin.

Considering first the trivial situation in which the source in Fig.3 is the identity, we can write the detection event operators in terms of the input event operators as

$$\bar{a}_{m1} = \int dk\, G(k)\, e^{ik(-x_{i1} - 2M\ln(x_{i1}) - t_i + \phi_1^-)}$$
$$\times \int d\Omega\, J(\Omega)\, e^{i\Omega(t_{d1} - \tau_1(t_i))}\bar{a}_{1,k,\Omega}$$

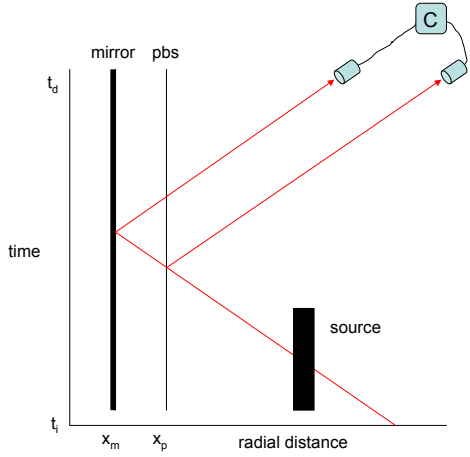

FIG. 3: Schematic of generic correlation experiment.

$$\bar{a}_{m2} = \int dk\, G(k)\, e^{ik(-x_{i2}-2Mln(x_{i2})-t_i+\phi_2^-)}$$
$$\times \int d\Omega\, J(\Omega)\, e^{i\Omega(t_{d2}-\tau_1(t_i))}\bar{a}_{2,k,\Omega} \quad (36)$$

From the continuity conditions at the mirror and polarizing beamsplitter we have

$$\begin{aligned}
\phi_1^- &= 2x_m + 4Mln(x_m) + (t_{d1} - x_{d1} - 2Mln(x_{d1})) \\
\phi_2^- &= 2x_p + 4Mln(x_p) + (t_{d2} - x_{d2} - 2Mln(x_{d2}))
\end{aligned} \quad (37)$$

where the boundary conditions at the detectors have been taken to be $\phi_1^+ = t_{d1} - x_{d1} - 2Mln(x_{d1})$ and $\phi_2^+ = t_{d2} - x_{d2} - 2Mln(x_{d2})$. Given these boundary conditions and recalling that the mode functions are invariant under free propagation, we can identify from Eqs 36 and 37 that $x_{i1} = -t_i + 2x_m + 4Mln(x_m) + t_{d1} - x_{d1} - 2Mln(x_{d1})$ and $x_{i2} = -t_i + 2x_p + 4Mln(x_p) + t_{d2} - x_{d2} - 2Mln(x_{d2})$ to be the points on the geodesic corresponding to the initial time $t_i$ and hence from Eq.35 find

$$\begin{aligned}
\tau_1(t_i) &= \int_{x_m}^{x_{d1}} \frac{dr'}{\sqrt{1-\frac{2M}{r'}}} + \int_{x_m}^{x_{i1}} \frac{dr'}{\sqrt{1-\frac{2M}{r'}}}. \\
&\approx -t_i + t_{d1} - Mln(\frac{x_{d1}x_{i1}}{x_m^2}) \\
\tau_2(t_i) &= \int_{x_p}^{x_{d2}} \frac{dr'}{\sqrt{1-\frac{2M}{r'}}} + \int_{x_p}^{x_{i2}} \frac{dr'}{\sqrt{1-\frac{2M}{r'}}}. \\
&\approx -t_i + t_{d2} - Mln(\frac{x_{d2}x_{i2}}{x_p^2}).
\end{aligned} \quad (38)$$

where we have simplified the results by assuming $r \gg 2M$ for all radii of interest.

### A. Classical Correlations

We can now include non-trivial source unitaries. We first consider classically correlated fields by considering equal displacements of the two polarization modes. From Eq.7 we get

$$\begin{aligned}
\bar{a}'_{m1} &= \bar{a}_{m1} + \alpha_1 \\
\bar{a}'_{m2} &= \bar{a}_{m2} + \alpha_2
\end{aligned} \quad (39)$$

where, from Eq.18, we have

$$\begin{aligned}
\alpha_1 &= \int dk\, |G(k)|^2\, e^{ik(\phi_1^- - \phi^c)}\alpha_{max} \\
\alpha_2 &= \int dk\, |G(k)|^2\, e^{ik(\phi_2^- - \phi^c)}\alpha_{max}.
\end{aligned} \quad (40)$$

The displacements are assumed to have the same spatial profile and have been correlated by setting the displacement phase, $\phi^c$, equal for both modes. The rate of coincidence detection, as analysed by the correlator, is given by

$$\begin{aligned}
C &= \langle 0|\bar{a}'^\dagger_{m1}\bar{a}'^\dagger_{m2}\bar{a}'_{m1}\bar{a}'_{m2}|0\rangle \\
&= |\alpha_1|^2|\alpha_2|^2
\end{aligned} \quad (41)$$

The operators annihilate when acting on the vacuum leaving the c-numbers as the only non-zero terms. The coincidence detection rate achieves its maximum value of $C = |\alpha_{max}|^4$ when $2x_m + 4Mln(x_m) + t_{d1} - x_{d1} - 2Mln(x_{d1}) = 2x_p + 4Mln(x_p) + t_{d2} - x_{d2} - 2Mln(x_{d2}) = \phi^c$. For simplicity consider the case of simultaneous detection, $t_{d1} = t_{d2}$. For flat space, $M = 0$ we have $2(x_p - x_m) = x_{d2} - x_{d1}$. That is, the extra path length traveled by the first mode between the polarizer and the mirror must be made up by placing an equivalent distance between the detectors. When the massive body is present the relation becomes $2(x_p - x_m + Mln(\frac{x_p^2 x_{d1}}{x_m^2 x_{d2}})) = x_{d2} - x_{d1}$. That is, the curvature now stretches space close to the body relative to far from the body such that the detectors must be moved further apart to observe maximum correlation. Notice that because the results depend only on the c-number displacements and these are unchanged by the generalization to event operators, so these results are identical to the standard approach. We conclude that in general classical correlations remain unchanged by the generalization to event operators.

### B. Non-classical Correlations

Now we consider the source in Fig.3 to be entangling. In particular we consider the production of time energy entanglement from vacuum inputs via the parametric unitary Eq.20. We obtain

$$\begin{aligned}
\bar{a}'_{m1} &= Cosh(\chi_{max})\bar{a}_{m1} + Sinh(\chi_1)\bar{a}^\dagger_{m2c} \\
\bar{a}'_{m2} &= Cosh(\chi_{max})\bar{a}_{m2} + Sinh(\chi_2)\bar{a}^\dagger_{m1c}
\end{aligned} \quad (42)$$

where

$$\begin{aligned}
\bar{a}_{m2c} &= \int dk\, G(k)\, e^{ik(-x_{i2}-2Mln(x_{i2})-t_i+\phi^c)} \\
&\quad \times \int d\Omega\, J(\Omega)\, e^{i\Omega(t_{d1}-\tau_1(t_i))}\bar{a}_{2,k,\Omega} \\
\bar{a}_{m1c} &= \int dk\, G(k)\, e^{ik(-x_{i1}-2Mln(x_{i1})-t_i+\phi^c)} \\
&\quad \times \int d\Omega\, J(\Omega)\, e^{i\Omega(t_{d2}-\tau_2(t_i))}\bar{a}_{1,k,\Omega}
\end{aligned} \tag{43}$$

For simplicity we will consider the case of weak parametric amplification for which $Cosh(\chi) \approx 1$ and $Sinh(\chi) \approx \chi$. Under this condition the rate of coincidence detection is given by

$$\begin{aligned}
C &= |\chi_2[\bar{a}_{m1}, \bar{a}_{m1c}^\dagger]|^2 \\
&= |\chi_2|^2 \int\int dk d\Omega |G(k)|^2 e^{ik(\phi_1^- - \phi^c)}|J(\Omega)|^2 e^{\Omega(\Delta)}
\end{aligned} \tag{44}$$

where

$$\begin{aligned}
\Delta &= Mln\left(\frac{x_{d1}x_{i1}x_p^2}{x_{d2}x_{i2}x_m^2}\right) \\
&\approx 2Mln\left(\frac{x_p}{x_m}\right)
\end{aligned} \tag{45}$$

and the approximation uses the assumption that the source and detectors are far away from the massive body.

If we first consider flat space, $M = 0$, then $\Delta = 0$ and the $\Omega$ integral will equal unity. Hence the coincidence count will depend only on the modal functions. As for the case of classical correlations we find maximum coincidence rate of $|\chi_{max}|^2$ occur when the detectors are positioned such that $2(x_p - x_m) = x_{d2} - x_{d1}$ (with $t_{d1} = t_{d2}$). When correctly positioned and timed the single detector rates are also both $|\chi_{max}|^2$, indicating perfect correlation. Again the event operator description agrees with the standard approach.

However when we consider the case $M \neq 0$ we find $\Delta \neq 0$ except for the trivial case in which there is no gap between the mirror and the polarizer ($x_m = x_p$). The detector position for the maximum coincidence rate is determined by the modal functions to occur (as for the classical case) when $2(x_p - x_m + 2Mln(\frac{x_p}{x_m})) = x_{d2} - x_{d1}$ (again with $t_{d1} = t_{d2}$ and assuming the detectors are far from the massive body). However the size of the maximum is reduced in the event operator formalism. In the limit that $\Delta \gg 1/\sigma_J$, where $\sigma_J$ is the variance of the distribution $J(\Omega)$ the coincidences will disappear to first order in $\chi$. Note though that the maximum single detector count rates remain $|\chi_{max}|^2$. Thus the effect of the different local propagation times in the event formalism is to decorrelate the entanglement.

To estimate the size of this effect we consider placing the source and detectors on a geostationary satellite with the mirror at ground level and the polarizing beamsplitter at height $h$. At geostationary orbit the curvature can be neglected and we find approximately

$$\Delta \approx 2M\frac{h}{r_e}. \tag{46}$$

We assume a Gaussian form for the function $J(\Omega)$,

$$J(\Omega) = \frac{d_t}{\sqrt{\pi}}e^{-\Omega^2 d_t^2} \tag{47}$$

As commented earlier, the effect of the $J(\Omega)$ function is to isolate a localized detection event that is then projected back onto the initial state. It seems natural then to associate $d_t$ with the temporal uncertainty in the measurement. Given that the detectors have been positioned to maximize the modal functions then the correlation function becomes

$$C = |\chi_{max}|^2 e^{-\frac{\Delta^2}{4d_t^2}} \tag{48}$$

and we conclude significant decorrelation will occur when $\Delta > 2d_t$. We estimate the *intrinsic* temporal uncertainty of a silicon photon counter to be around 200 fs and hence set the standard deviation in units of length to $d_t = 6 \times 10^{-5}m$. Using Eq.46, the mass of earth in units of length, $M = 4.4 \times 10^{-3}m$ and the radius of earth $r_e = 6.38 \times 10^6 m$ we find this implies significant decorrelation when $h > 90km$.

### C. An Experimental Proposal

The estimate at the close of the last section suggests that a testable effect exists for Earth scale curvatures. None-the-less, directing entangled beams down from geostationary orbit to reflectors separated by hundreds of kilometers and back is not currently practical. However a slight rearrangement of the set-up, shown in Fig.4, leads to a more practical proposal. We now assume that the source, polarizing beamsplitter and second detector are all approximately at height $x_p = r_e + h$, whilst the mirror, first detector and the correlator are all approximately at ground-level, $x_m = r_e$. A classical channel links the second detector and the correlator. Mathematically the situation is still described by the general equations of the previous section. In particular it is still possible to maximize the modal correlation function, though clearly we must now allow for different detection times. The first line of Eq.45 still describes the magnitude of $\Delta$ but now with $x_{d1} \approx x_m$ and $x_{d2} \approx x_p$. With the modal functions maximized (which implies $x_{i1} = x_{i2}$) we have

$$\Delta \approx Mln\left(\frac{x_p}{x_m}\right) \tag{49}$$

Following the arguments of the previous section we thus conclude that the correlations between detection of one beam of a parametric source on a satellite and the subsequent detection of the other beam at ground level will be significantly reduced when $h > 180km$.

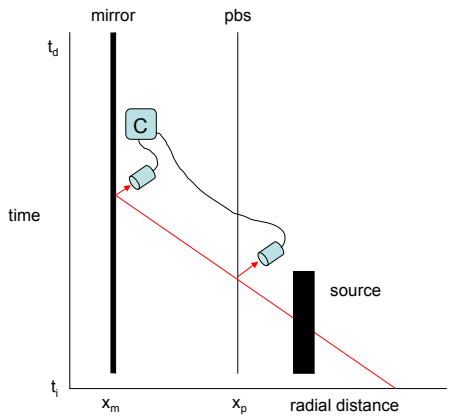

FIG. 4: Schematic of modified correlation experiment. Now the source, polarizing beamsplitter and second detector are approximately at height $x_p$, whilst the mirror, first detector and the correlator are approximately at height $x_m$. A classical communication channel sends the information from the second detector to the correlator.

## V. CONCLUSION

Motivated by toy models of exotic general relativistic potentials and more general considerations we have introduced a non-standard formalism for analyzing quantum optical fields on a curved background metric. In contrast to the standard approach in terms of global mode operators, our non-standard formalism involves local event operators that act on Hilbert sub-spaces that are localized in space-time. As such the quantum connectivity of space-time is reduced in our model. We have shown that for inertial observers in a flat space-time the predictions of the standard and non-standard formalisms agree. However, for entangled states in curved space-times differences can arise. To illustrate this we have studied the effect on optical entanglement of evolution through varying gravitational fields using both formalisms. The new formalism predicts a decorrelation effect that could be observable under experimentally achievable conditions.

The novelty of this new predicted effect should not be underestimated. Although previous studies have found decorrelation of entanglement in non-inertial frames [14] the effects are much smaller than the one predicted here. They also differ from the ones found here in several ways. First note that although, because of the loss of photon correlations, one might refer to this effect as decoherence, in fact the effect is in principle reversible. Considering the set-up of Fig.3, correlation would be regained by re-sending (before detection) mode 1 along mode 2's path and vice versa. Secondly we anticipate that more unusual evolutions may arise for strongly entangled qubit states as suggested in Ref [15]. Treatment of such situations with the same rigour as used here would require consideration of highly non-linear Heisenberg evolutions that are beyond the scope of the present calculations.

We believe that an experimental investigation of this predicted effect could be warranted, for if observed, it would represent a new phenomenon with major consequences for quantum physics in general and quantum information in particular.

This work was supported by the Defence Science and Technology Organization and the Australian Research Council.

[1] N.D.Birrell and P.C.W.Davies, *Quantum fields in curved space*, (Cambridge University Press 1982).
[2] S. W. Hawking, *Commun.Math.Phys.* **43**, 199 (1975); R. Gambini, R. A. Porto and J. Pullin, *Phys.Rev.Lett* **93**, 240401 (2004).
[3] M.S.Morris, K.P.Thorne and U.Yurtsever, Phys.Rev.Lett. **61**, 1446 (1988).
[4] D. Deutsch, Phys.Rev.D **44**, 3197 (1991).
[5] D.Bacon, Phys.Rev.A, **70** 032309 (2004).
[6] T.C.Ralph, Phys. Rev. A **76**, 012336 (2007).
[7] D. Bouwmeester et al., in N. Dadhich and J. Narlikar (Eds.), Gravitation and Relativity: At the turn of the Millennium (IUCAA, Pune, 1998).
[8] S. Carlip, *Is Quantum Gravity Necessary?*, arXiv:0803.3456 (2008).
[9] P.T. Cochrane, G.J. Milburn, W.J. Munro, Phys. Rev. A **62**, 062307 (2000).
[10] Between them Eqs 10 and 12 represent the so-called positive frequency components of the field. Their conjugates represent the negative frequency components.
[11] W.G.Grice, A.B.U'ren, I.A.Walmsley, Phys. Rev. A **64** 063815 (2001).
[12] R.M.Wald, *General Relativity* (University of Chicago Press. 1984).
[13] E. F. Taylor and J. A. Wheeler, *Exploring Black Holes; Introduction to General Relativity* (Addison Wesley Longman, San Francisco, 2000).
[14] P. M. Alsing and G. J. Milburn, *Phys.Rev.Lett*, **91** 180404 (2003); I. Fuentes-Schuller and R. B. Mann, *Phys.Rev.Lett.* **95**, 120404 (2005).
[15] T. C. Ralph, *Proc. SPIE* **6305**, 63050P (2006).